\documentclass[preprint]{aastex}

\usepackage{multirow}
\usepackage{graphicx}
\usepackage{amssymb}
\usepackage{bigdelim}
\usepackage{natbib}
\usepackage{array}
\usepackage{amsmath}
\usepackage{rotating}
\usepackage{subfigure}
\usepackage{color}

\usepackage[pdftitle={wasp35-48-51}]{hyperref}
\hypersetup{pdftitle=wasp35-48-51}

\bibpunct{(}{)}{;}{a}{}{,}

\newcommand{\teff}{\mbox{$T_{\rm eff}$}}
\newcommand{\logg}{\mbox{$\log g$}}
\newcommand{\vsini}{\mbox{$v \sin i$}}
\newcommand{\mictrb}{\mbox{$\xi_{\rm t}$}}
\newcommand{\mactrb}{\mbox{$v_{\rm mac}$}}

\newcommand{\kms}{\mbox{km\,s$^{-1}$}}

\shorttitle{WASP-35b, -48b and -51b}
\shortauthors{B.Enoch et al.}

\begin{document}

\title{WASP-35b, WASP-48b and WASP-51b: Two new planets and an independent discovery of HAT-P-30b.}

\author{B.Enoch\altaffilmark{1}, D.R.Anderson\altaffilmark{2}, S.C.C.Barros\altaffilmark{3}, D.J.A.Brown\altaffilmark{1}, A.Collier Cameron\altaffilmark{1}, F.Faedi\altaffilmark{3}, M.Gillon\altaffilmark{4}, G.H\'ebrard\altaffilmark{5,6}, T.A.Lister\altaffilmark{7}, D.Queloz\altaffilmark{8}, A.Santerne\altaffilmark{9}, B.Smalley\altaffilmark{2}, R.A.Street\altaffilmark{7}, A.H.M.J.Triaud\altaffilmark{8}, R.G.West\altaffilmark{10}, F.Bouchy\altaffilmark{5,6}, J.Bento\altaffilmark{11}, O.Butters\altaffilmark{10}, L.Fossati\altaffilmark{12}, C.A.Haswell\altaffilmark{12}, C.Hellier\altaffilmark{2}, S.Holmes\altaffilmark{12}, E.Jehen\altaffilmark{4}, M.Lendl\altaffilmark{8}, P.F.L.Maxted\altaffilmark{2}, J.McCormac\altaffilmark{3}, G.R.M.Miller\altaffilmark{1}, V.Moulds\altaffilmark{3}, C.Moutou\altaffilmark{9}, A.J.Norton\altaffilmark{12}, N.Parley\altaffilmark{1}, F.Pepe\altaffilmark{8}, D.Pollacco\altaffilmark{3}, D.Segransan\altaffilmark{8}, E.Simpson\altaffilmark{3}, I.Skillen\altaffilmark{13}, A.M.S.Smith\altaffilmark{2}, S.Udry\altaffilmark{8} and P.J.Wheatley\altaffilmark{11}
}

\altaffiltext{1}{SUPA, School of Physics and Astronomy, University of St. Andrews, North Haugh, St Andrews, KY16 9SS.}
\altaffiltext{2}{Astrophysics Group, Keele University, Staffordshire, ST5 5BG, UK.}
\altaffiltext{3}{Astrophysics Research Centre, School of Mathematics \& Physics, Queen's University Belfast, University Road, Belfast, BT7 1NN, UK.}
\altaffiltext{4}{Institut d`Astrophysique et de G\'eophysique, Universit\'e de Li\'ege, All\'ee de 6 Ao\^ut, 17, Bat B5C, Li\'ege 1, Belgium.}
\altaffiltext{5}{Institut d'Astrophysique de Paris, UMR7095 CNRS, Universit\'e Pierre \& Marie Curie, France.}
\altaffiltext{6}{Observatoire de Haute-Provence, CNRS/OAMP, 04870 St Michel l'Observatoire, France}
\altaffiltext{7}{Las Cumbres Observatory, 6740 Cortona Drive Suite 102, Goleta, CA 93117, USA.}
\altaffiltext{8}{Observatoire astronomique de l`Universit\'e de Gen\'eve, 51 Chemin des Maillettes, 1290 Sauverny, Switzerland.}
\altaffiltext{9}{Laboratoire d'Astrophysique de Marseille, Universit\'e d'Aix-Marseille \& CNRS, 38 rue Fr\'ed\'eric Joliot-Curie, 13388 Marseille cedex 13, France}
\altaffiltext{10}{Department of Physics and Astronomy, University of Leicester, Leicester, LE1 7RH, UK.}
\altaffiltext{11}{Department of Physics, University of Warwick, Coventry, CV4 7AL, UK}
\altaffiltext{12}{Department of Physics and Astronomy, The Open University, Milton Keynes, MK7 6AA, UK}
\altaffiltext{13}{Isaac Newton Group of Telescopes, Apartado de Correos 321, E-38700 Santa Cruz de Palma, Spain}

\begin{abstract}
We report the detection of WASP-35b, a planet transiting a metal-poor ([Fe/H]~=~$-0.15$) star in the Southern hemisphere, WASP-48b, an inflated planet which may have spun-up its slightly evolved host star of $1.75 R_{\odot}$ in the Northern hemisphere, and the independent discovery of HAT-P-30b / WASP-51b, a new planet in the Northern hemisphere. Using WASP, RISE, FTS and TRAPPIST photometry, with CORALIE, SOPHIE and NOT spectroscopy, we determine that WASP-35b has a mass of $0.72\pm0.06 M_J$ and radius of $1.32\pm0.03 R_J$, and orbits with a period of 3.16~days, WASP-48b has a mass of $0.98\pm0.09 M_J$, radius of $1.67\pm0.08 R_J$ and orbits in 2.14~days, while WASP-51b, with an orbital period of 2.81~days, is found to have a mass of $0.76\pm0.05 M_J$ and radius of $1.42\pm0.04 R_J$, agreeing with values of $0.71\pm0.03 M_J$ and $1.34\pm0.07 R_J$ reported for HAT-P-30b.
\end{abstract}

\maketitle

\section{Introduction}
\label{s:intro}


Around 120 transiting exoplanets have now been discovered\footnotemark \footnotetext[1]{www.exoplanet.eu}, having both their radii and masses known, and may be subjected to further investigations into their bulk and atmospheric composition \citep[see e.g.][]{char02,char05,fortney07,haswell10,seager10}. They show a diverse range of densities and thus internal compositions. Over 80\% of those with masses greater than $0.1 M_J$ have radii greater than that of Jupiter, and many have extremely low densities which even coreless models struggle to explain \citep{fortney07, burrows07}, for example TrES-4b \citep{mandushev07}, WASP-17b \citep{anderson10, anderson11} and Kepler-7b \citep{latham10}. More transiting exoplanets are vital to constrain the formation and evolution of such planets. 

Several possible effects on planetary radii have been discussed in the literature, including inflation due to strong heating by irradiation received from the host star \citep{guillot96, guillot02}, ohmic heating from the coupling of magnetic fields and atmospheric flows \citep{batygin10,batygin11} and tidal heating due to the circularisation of close-in exoplanets \citep{bodenheimer01, bodenheimer03, jackson08}.

Most of the known transiting exoplanets orbit very close to their host star, which produces strong tidal forces between them. The tidal interactions may result in orbital circularisation, synchronisation and decay \citep{pont09}. An apparent relationship between period and eccentricity of non-transiting exoplanets, such that planets very close to their host stars have circular orbits while those farther from the stars show a wide range of eccentricities, seems to provide evidence for the tidal circularisation of planetary orbits \citep{mazeh08}. However, it had generally been assumed that the masses of planets are too small to synchronise the stellar rotation with the planetary orbit through tidal forces \citep{mazeh08}. Some possible exceptions exist for planets with large masses, for example the $\tau$~Boo system with a $4 M_J$ planet, for which there is evidence for stellar rotation synchronisation \citep{fares09}. Recently, CoRoT-11b has been announced to have a `peculiar tidal evolution' \citep{gandolfi11}, where the stellar rotation period of 1.7~days is now shorter than the planetary orbital period of 3~days, due to the strong tidal forces causing orbital decay of the planet from a previously approximately synchronised orbital period. The possibility of perhaps unknown close-in exoplanets affecting the stellar spin rate of their host stars produces uncertainty in the use of the gyrochronological age estimate of a star, based on the normal spin down of a star as it ages. 
 
One motivation of the WASP project is to discover enough transiting exoplanets with a wide range of orbital and compositional parameters to allow analyses that may distinguish between different models. Recent WASP discoveries help fill in areas of planetary mass-radius paramter space, for example WASP-39b \citep{faedi11} is the least dense sub-Saturn mass planet known. 

Here, we announce WASP-35b, WASP-48b and WASP-51b ($\equiv$~HAT-P-30): the positions and magnitudes of each star are given in Table \ref{tab:loc}. We set out the photometric and spectroscopic observations obtained for each system, before presenting the resulting stellar and system parameters. WASP-48 appears to be a slightly evolved star with a large radius of $1.75 R_{\odot}$ and a very short rotation period of around 7~days for its age, around 8~Gyr. WASP-48b may therefore have spun-up its host star despite a mass of only around $1 M_J$, due to the star's inflated radius, although the system has not yet reached synchronisation of stellar rotation period with planetary orbital period (2.15~days). As the host star continues to evolve, increasing significantly in radius, it may engulf WASP-48b, which is currently orbiting at a distance of around 4 times the stellar radius, unless orbital decay due to tidal interactions, such as may have happened to CoRoT-11b \citep{gandolfi11}, push the planet far enough away to survive.

\begin{table}[h!]
\caption{Star locations.}
\label{tab:loc}
\begin{center}
\begin{tabular}{cccc}
\hline
Star & RA (J2000) & Dec (J2000) & V mag (NOMAD) \\
\hline
WASP-35 & 05 04 19.6 & $-$06 13 47 & 10.95 \\
WASP-48 & 19 24 39.0 & +55 28 23 & 11.66 \\
WASP-51 & 08 15 48.0 & +05 50 12 & 10.36 \\
\hline
\end{tabular}
\end{center}
\end{table}


\section{Observations}
\label{obs}

\subsection{Photometric Obervations}

The WASP-North and South observatories are located at La Palma, Canary Islands and SAAO, South Africa respectively, and each consist of eight 11cm telescopes of $7.8^{\circ} \times 7.8^{\circ}$ field of view each, on a single fork mount. The cameras scan repeatedly through eight to ten sets of fields each night, taking 30 second exposures. See \citet{pollacco06} for more details on the WASP project and the data reduction procedure, and \citet{cameron07} and \citet{pollacco08} for an explanation of the candidate selection process.

High precision transit light curves of WASP-35, WASP-48 and WASP-51 were obtained with the RISE instrument mounted in the 2.0m Liverpool Telescope \citep{steele08, gibson08} at the Observatorio del Roque de los Muchachos on La Palma, Canary Islands. The data were reduced with the ULTRACAM pipeline \citep{dhillon07} following the same procedure as \citet{barros11}. Further high quality photometry of WASP-35 was obtained with the 0.6m TRAnsiting Planets and PlanetesImals Small Telescope (TRAPPIST) at La Silla Observatory, Chile, and with the LCOGT 2.0m Faulkes Telescope South (FTS) at Siding Spring Observatory. 

\begin{figure}
\begin{center}
\includegraphics[angle=0,width=100mm]{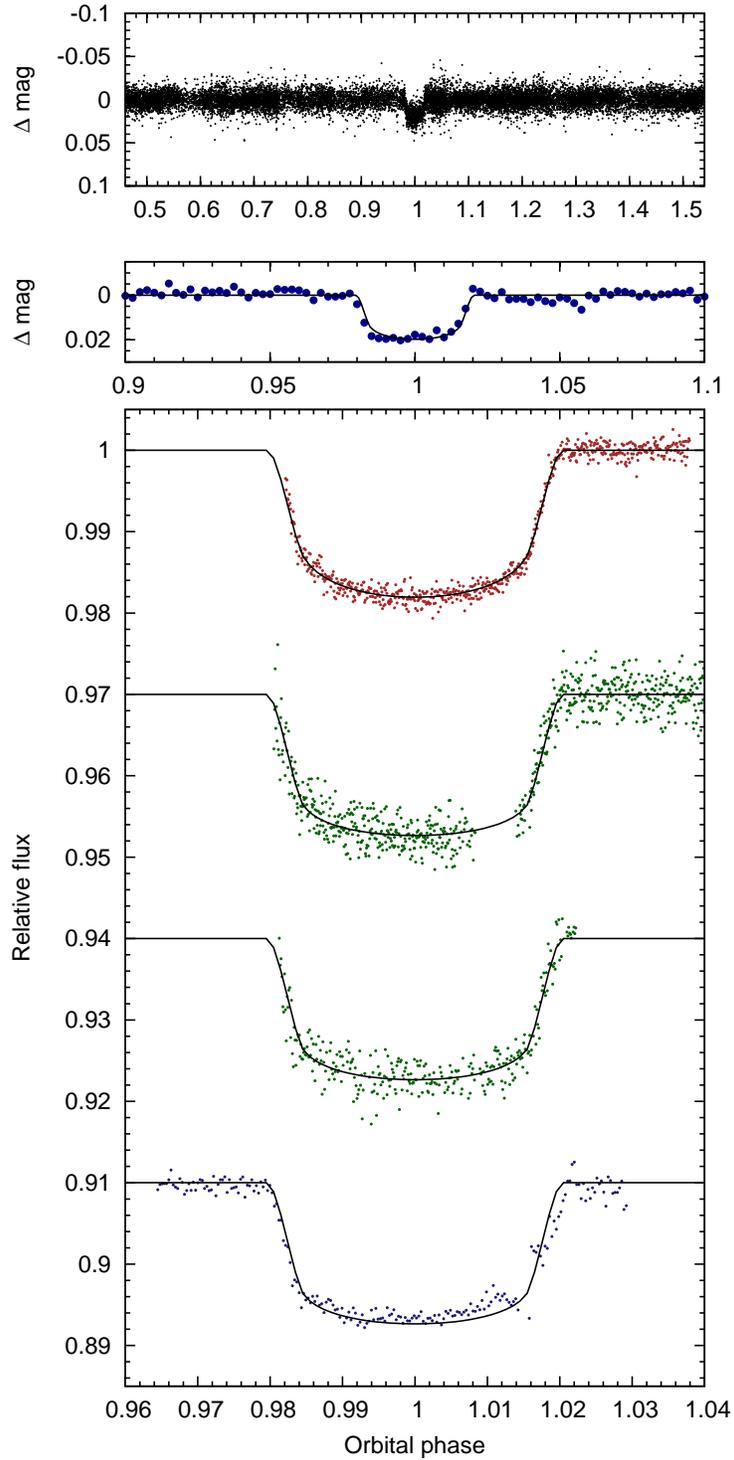}
\caption{Upper plot: WASP-35 discovery lightcurve folded on the orbital period of $P$~=~3.16~d. Middle plot: WASP data binned in phase, with bin-width = 11~minutes. Lower plot (top to bottom): transit obtained with the RISE instrument on 30 November 2010, TRAPPIST on 23 December 2010 and 11 January 2011, and FTS on 1 January 2011, lightcurves offset for clarity.}
\label{fig:wasp35phot}
\end{center}
\end{figure}

\begin{figure}
\begin{center}
\includegraphics[angle=0,width=125mm]{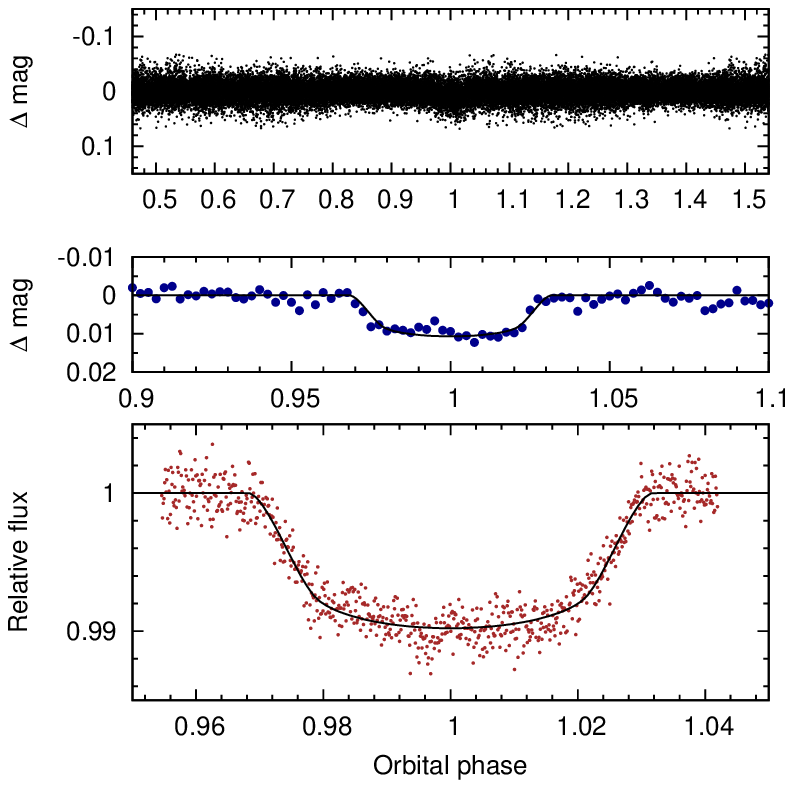}
\caption{Upper plot: WASP-48 discovery lightcurve folded on the orbital period of $P$~=~2.14~d. Middle plot: WASP data binned in phase, with bin-width = 8~minutes. Lower plot: transit obtained with the RISE instrument on 1 July 2010.}
\label{fig:wasp48phot}
\end{center}
\end{figure}

\begin{figure}
\begin{center}
\includegraphics[angle=0,width=125mm]{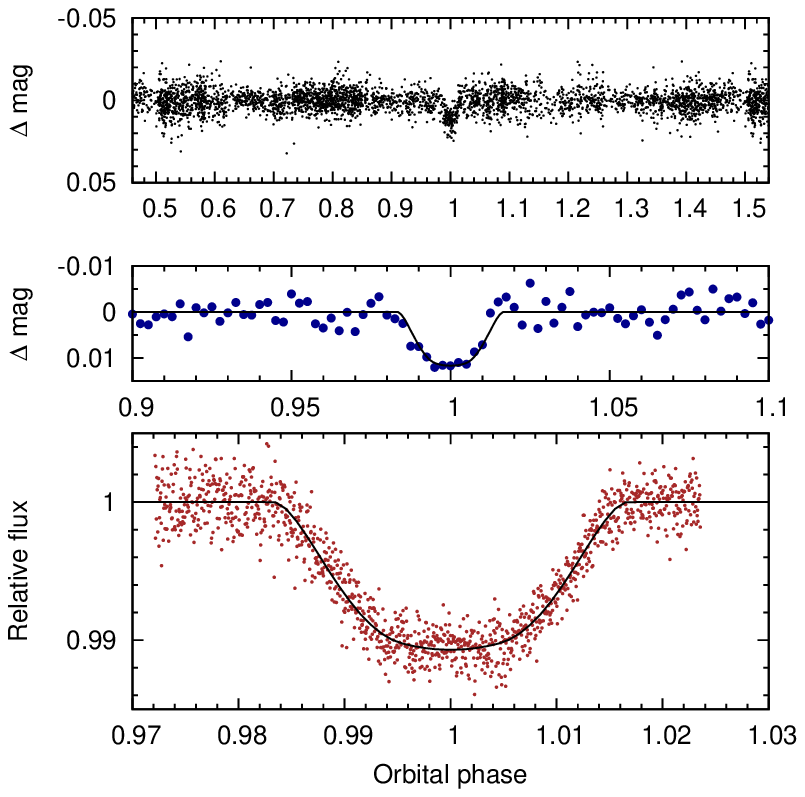}
\caption{Upper plot: WASP-51 discovery lightcurve folded on the orbital period of $P$~=~2.81~d. Middle plot: WASP data binned in phase, with bin-width = 10~minutes. Lower plot: transit obtained with the RISE instrument on 12 January 2011.}
\label{fig:wasp51phot}
\end{center}
\end{figure}

Photometric observations for each of WASP-35, WASP-48 and WASP-51 are detailed in Table \ref{tab:photobs}, and shown in Figures \ref{fig:wasp35phot} to \ref{fig:wasp51phot}.

\begin{table}[h!]
\caption{Photometric Observations}
\label{tab:photobs}
\begin{center}
\begin{tabular}{|l|r|r|}
\hline
Instrument & Date(s) & Num datapoints \\
\hline
\hline
\multicolumn{3}{c}{ } \\
\multicolumn{3}{c}{WASP-35} \\
\hline
\multirow{2}{*}{WASP-N (2 cams)} & 10/2008-01/2009 &  \multirow{3}{*}{16,748} \\
                                          & 10/2009-01/2010 & \\
        WASP-S (1 cam)  & 10/2009-01/2010 & \\
\hline
       RISE (V+R) & 30/11/2010 & 630 \\
\hline
       TRAPPIST (I+z) & 23/12/2010 & 742 \\
\hline
       FTS (z) & 01/01/2011 & 205 \\
\hline
       TRAPPIST (I+z) & 11/01/2011 & 300 \\
\hline
 \multicolumn{3}{c}{ } \\
\multicolumn{3}{c}{WASP-48} \\
\hline
 \multirow{4}{*}{WASP-N (3 cams)} & 05/2007-07/2007 & \multirow{4}{*}{56,036} \\
                                      & 05/2008-08/2008  & \\
                                      & 05/2009-06/2009  & \\
                                      & 05/2010-09/2010  & \\
\hline
         RISE (V+R) & 01/07/2010 & 770 \\
\hline
 \multicolumn{3}{c}{ } \\
\multicolumn{3}{c}{WASP-51} \\
\hline
 WASP-N (1 cam) & 11/2009-03/2010 & 4,453 \\
\hline
         RISE (V+R) & 12/01/2011 & 1,245 \\
\hline
\end{tabular}
\end{center}
\end{table}


\subsection{Spectroscopic Observations}

We obtained 3 radial velocity measurements of WASP-35 on the 25 and 26 January 2010 with the FIbre-fed Echelle Spectrograph (FIES) on the 2.5m Nordic Optical Telescope\footnotemark \footnotetext[2]{operated on the island of La Palma jointly by Denmark, Finland, Iceland, Norway and Sweden, in the Spanish Observatorio del Roque de los Muchachos of the Instituto de Astrofisica de Canarias}. FIES was used in medium resolution mode (R=46,000) with simultaneous ThAr calibration, and the observations were reduced using the FIEStool package\footnotemark \footnotetext[3]{http://www.not.iac.es/instruments/fies/fiestool/FIEStool.html}, then cross-correlated with a high signal-to-noise spectrum of the Sun to obtain the radial velocities. A further nine datapoints were obtained with the CORALIE echelle spectrograph mounted on the Swiss telescope at the ESO La Silla Observatory in Chile on 5 January 2010 and between 6 and 14 February 2010 and the resulting spectra were processed with the standard data reduction CORALIE pipeline \citep{baranne96,mayor09}, plus a correction for the blaze function \citep{triaud10}. 

WASP-48 was observed spectroscopically with the SOPHIE fibre-fed echelle spectrograph mounted on the 1.93m telescope at Observatoire de Haute-Provence (OHP) \citep{perruchot08,bouchy09}. Fourteen datapoints were taken in High Efficienty mode (R~=~40,000) between 19 April and 18 October 2010, obtaining a signal-to-noise ratio of around 35 for each. Wavelength calibration with a Thorium-Argon lamp was performed to allow the interpolation of the spectral drift of SOPHIE.

Radial velocity observations of WASP-51 were made with the CORALIE and SOPHIE spectrographs. Nine datapoints were taken with SOPHIE between 16 October and 25 November 2010, using the high resolution mode (R~=~75,000) and obtaining a signal-to-noise ratio of around 20 for each measurement. Ten further datapoints were obtained with CORALIE between 7 December 2010 and 4 January 2011. 

The CORALIE guiding camera revealed a faint star located at about 1.5\arcsec\ from WASP-51 which was observed simultaneously with WASP-51 through the 3\arcsec\ fiber of the SOPHIE spectrograph. Thus, we performed radial velocity blend simulations (Santerne et al, in preparation) in order to investigate whether small bisector variations seen in SOPHIE data are due to this secondary blending star, though the bisector spans are compatible with no variation within the $2\sigma$ level. This simulation consisted of producing SOPHIE cross-correlation functions (CCFs) for WASP-51 with the unblended orbital solution seen by CORALIE and blending a single solar star with different systemic radial velocity, distance from WASP-51 and \vsini. With 10,000 Monte Carlo simulations in the ranges 34.6-54.6~\kms\ for the systemic RV, 1-5~times the distance of WASP-51 and 4-40~\kms\ for the \vsini\ of the blending star, we didn't find any significant model that reproduces the bisector effect seen in the SOPHIE data. For the case of WASP-51, the effect of the blending star on the bisector is less than a few m.s$^{-1}$. We concluded that the faint star 1.5\arcsec\ away didn't affect the SOPHIE CCF of WASP-51.

\begin{figure}
\begin{center}
\includegraphics[width=125mm]{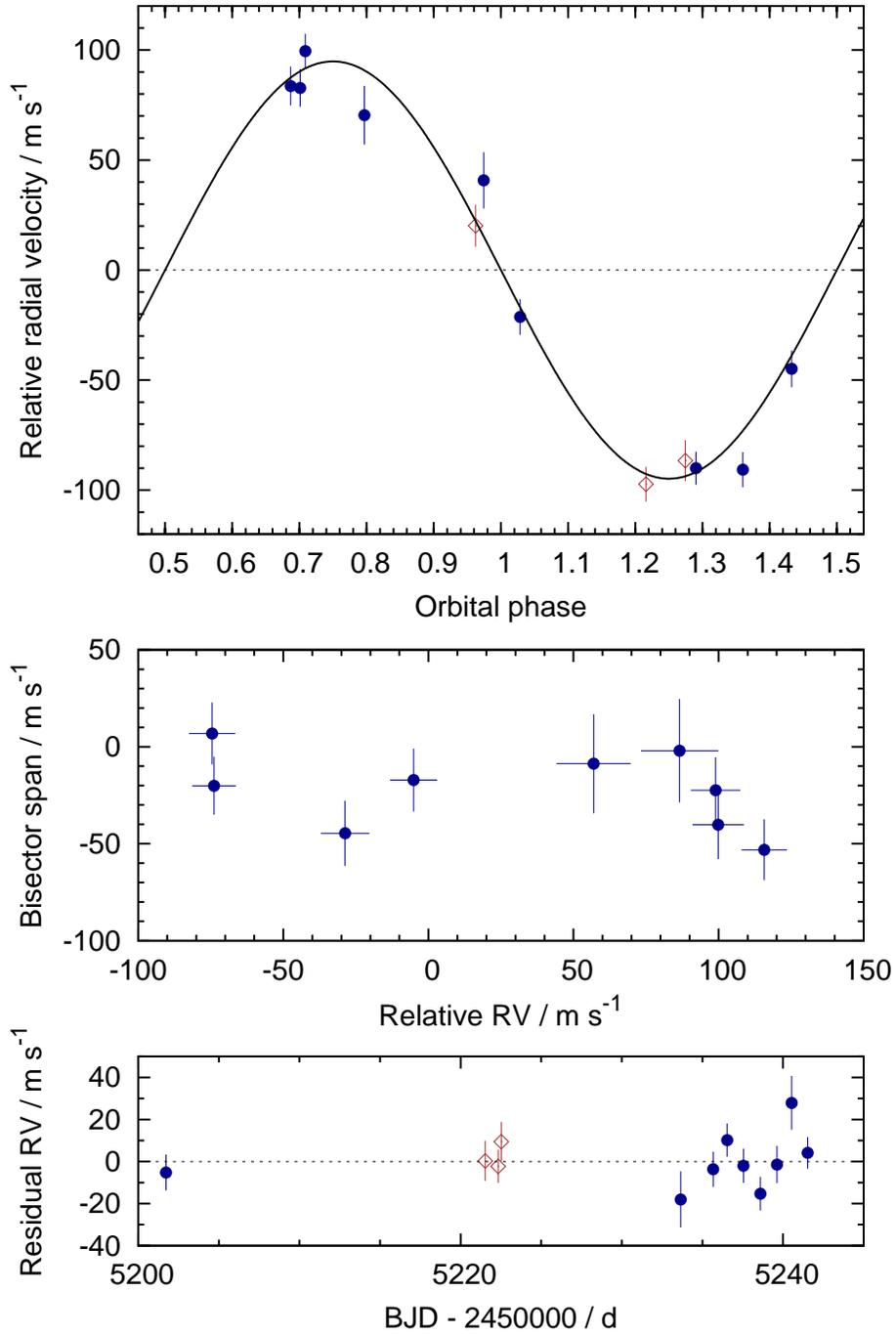}
\caption{Radial velocity measurements for WASP-35: red, open diamonds denote NOT observations, blue circles are from CORALIE. The solid line shown on the top plot is the best-fitting MCMC solution with eccentricity fixed to zero. The centre-of-mass velocity, $\gamma$ = 17.71 km s$^{-1}$, was subtracted. The middle plot shows the bisector span variations against radial velocity measurements. The uncertainty in the bisector span measurements are taken to be twice the uncertainty in the radial velocity measurements. The lower plot shows the radial velocity residuals against time.}
\label{fig:rv35}
\end{center}
\end{figure}

\begin{figure}
\begin{center}
\includegraphics[width=125mm]{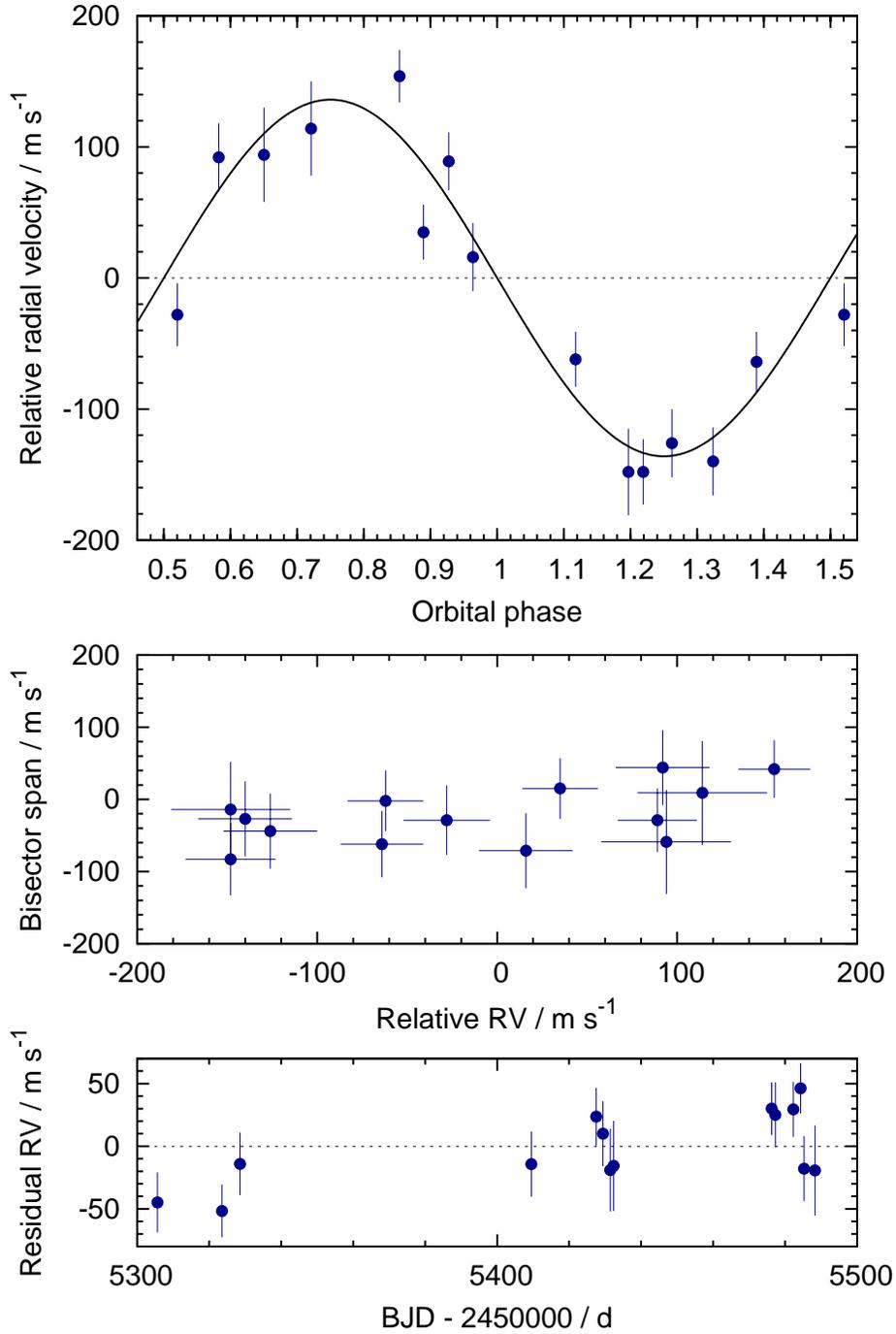}
\caption{Radial velocity measurements for WASP-48: details are as for WASP-35, with centre-of-mass velocity, $\gamma$ = -19.68 km s$^{-1}$ here.}
\label{fig:rv48}
\end{center}
\end{figure}

\begin{figure}
\begin{center}
\includegraphics[width=125mm]{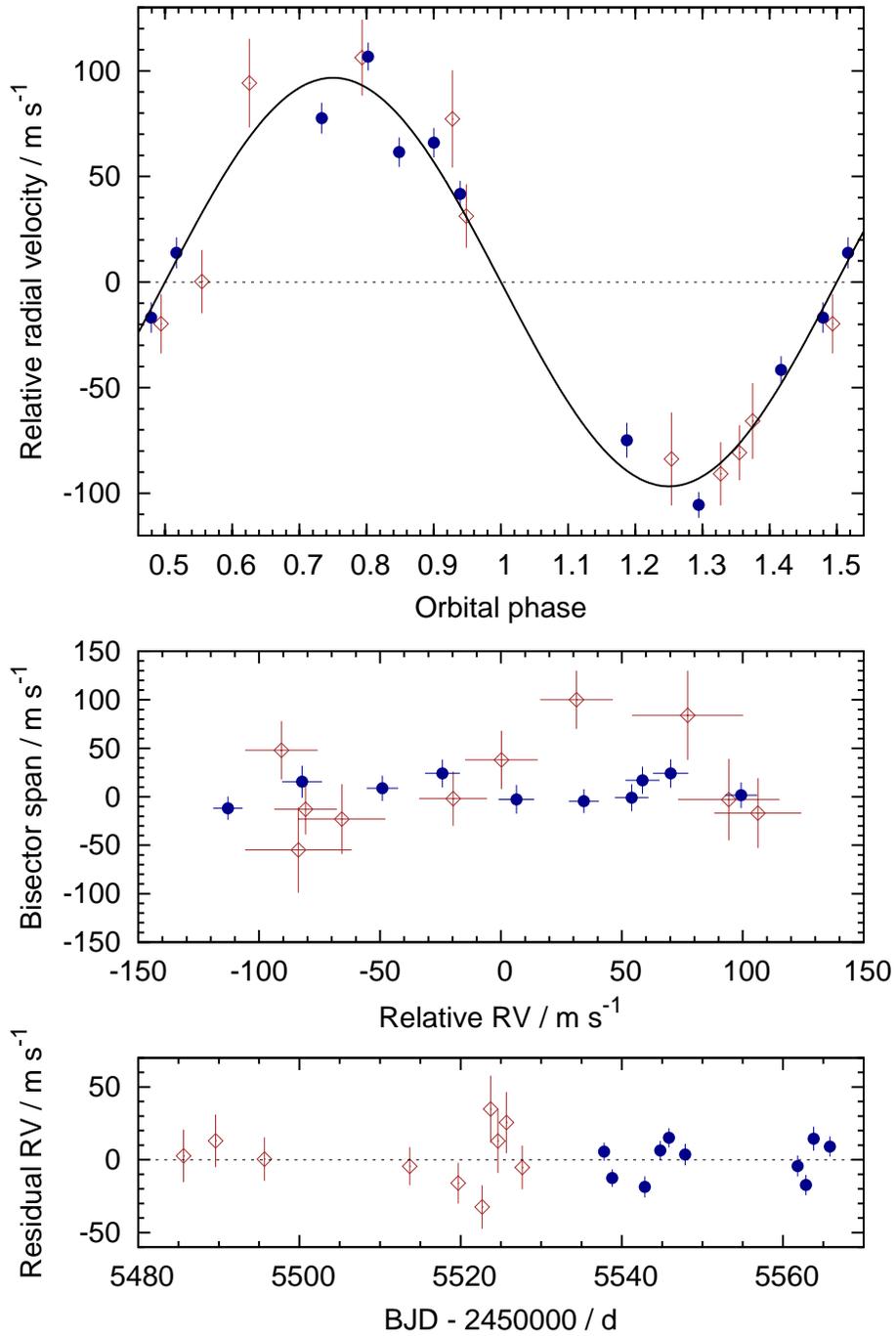}
\caption{Radial velocity measurements for WASP-51: red diamonds and blue circles denote OHP and CORALIE observations, respectively. Details are as for WASP-35, with centrre-of-mass velocity, $\gamma$ = 44.68 km s$^{-1}$ here.}
\label{fig:rv51}
\end{center}
\end{figure}

The radial velocities measured using these spectra are given in Tables \ref{tab:rv35}, \ref{tab:rv48} and \ref{tab:rv51}, and shown as a function of orbital phase in Figures \ref{fig:rv35}, \ref{fig:rv48} and \ref{fig:rv51}, along with radial velocity residuals and bisector span variations. The radial velocity measurements show low-amplitude variations, compatible with the existence of orbiting planets, and are not correlated with the bisector spans (available for CORALIE and SOPHIE data) \citep{queloz01}, which rules out stellar blend configurations.

\begin{table}
\caption{Radial velocity measurements of WASP-35}
\label{tab:rv35}
\begin{center}
\begin{tabular}{ccccl}
\hline
BJD--2\,400\,000 & RV & $\sigma$$_{\rm RV}$ & BS & Instrument \\
 & (km s$^{-1}$) & (km s$^{-1}$) & (km s$^{-1}$) & \\
\hline
5201.7298 & 17.8174 & 0.0085 & $-$0.0225 & CORALIE \\
5221.5256 & 17.7386 & 0.0096 & & FIES \\
5222.3276 & 17.6211 & 0.0079 & & FIES \\
5222.5131 & 17.6318 & 0.0093 & & FIES \\
5233.6480 & 17.8050 & 0.0133 & $-$0.0020 & CORALIE \\
5235.6597 & 17.6897 & 0.0084 & $-$0.0446 & CORALIE \\
5236.5322 & 17.8342 & 0.0078 & $-$0.0531 & CORALIE \\
5237.5430 & 17.7133 & 0.0081 & $-$0.0172 & CORALIE \\
5238.5912 & 17.6439 & 0.0080 & ~0.0069 & CORALIE \\
5239.6245 & 17.8183 & 0.0088 & $-$0.0402 & CORALIE \\
5240.5331 & 17.7754 & 0.0128 & $-$0.0087 & CORALIE \\
5241.5318 & 17.6446 & 0.0075 & $-$0.0201 & CORALIE \\
\hline
\end{tabular}
\end{center}
\end{table}

\begin{table}
\caption{Radial velocity measurements of WASP-48 taken with the SOPHIE spectrograph}
\label{tab:rv48}
\begin{center}
\begin{tabular}{cccc}
\hline
BJD--2\,400\,000 & RV & $\sigma$$_{\rm RV}$ & BS \\
 & (km s$^{-1}$) & (km s$^{-1}$) & (km s$^{-1}$) \\
\hline
5305.6441 & $-$19.711 & 0.024 & $-$0.029 \\
5323.5842 & $-$19.648 & 0.021 & ~0.015 \\
5328.5784 & $-$19.831 & 0.025 & $-$0.083 \\
5409.4890 & $-$19.667 & 0.026 & $-$0.071 \\
5427.5503 & $-$19.747 & 0.023 & $-$0.062 \\
5429.4215 & $-$19.809 & 0.026 & $-$0.044 \\
5431.4254 & $-$19.831 & 0.033 & $-$0.014 \\
5432.3967 & $-$19.589 & 0.036 & $-$0.059 \\
5476.2714 & $-$19.745 & 0.021 & $-$0.002 \\
5477.2679 & $-$19.591 & 0.026 & ~0.044 \\
5482.2945 & $-$19.594 & 0.022 & $-$0.029 \\
5484.2801 & $-$19.529 & 0.020 & ~0.042 \\
5485.2880 & $-$19.823 & 0.026 & $-$0.027 \\
5488.2831 & $-$19.569 & 0.036 & ~0.009 \\
\hline
\end{tabular}
\end{center}
\end{table}

\begin{table}
\caption{Radial velocity measurements of WASP-51}
\label{tab:rv51}
\begin{center}
\begin{tabular}{ccccl}
\hline
BJD--2\,400\,000 & RV & $\sigma$$_{\rm RV}$ & BS & Instrument \\
 & (km s$^{-1}$) & (km s$^{-1}$) & (km s$^{-1}$) & \\
\hline
5485.6249 & 44.611 & 0.018 & $-$0.023 & SOPHIE \\        
5489.6119 & 44.783 & 0.018 & $-$0.017 & SOPHIE \\       
5495.6696 & 44.708 & 0.015 & ~0.100 & SOPHIE \\        
5513.6757 & 44.596 & 0.013 & $-$0.013 & SOPHIE \\       
5519.6870 & 44.657 & 0.014 & $-$0.002 & SOPHIE \\       
5522.6691 & 44.677 & 0.015 & ~0.038 & SOPHIE \\       
5523.7176 & 44.754 & 0.023 & ~0.084 & SOPHIE \\       
5524.6337 & 44.593 & 0.022 & $-$0.055 & SOPHIE \\       
5525.6785 & 44.771 & 0.021 & $-$0.003 & SOPHIE \\       
5527.6506 & 44.586 & 0.015 & ~0.048 & SOPHIE \\
5537.8023 & 44.7111 & 0.0062 & $-$0.0046 & CORALIE \\
5538.8006 & 44.5639 & 0.0061 & $-$0.0118 & CORALIE \\
5542.8449 & 44.7469 & 0.0073 & ~0.0240 & CORALIE \\
5544.7669 & 44.6278 & 0.0065 & ~0.0087 & CORALIE \\
5545.8487 & 44.7761 & 0.0066 & ~0.0016 & CORALIE \\
5547.8576 & 44.6832 & 0.0074 & $-$0.0027 & CORALIE \\
5561.8052 & 44.6526 & 0.0072 & ~0.0241 & CORALIE \\
5562.8426 & 44.7309 & 0.0070 & $-$0.0010 & CORALIE \\
5563.7958 & 44.5945 & 0.0083 & ~0.0155 & CORALIE \\
5565.7984 & 44.7353 & 0.0070 & ~0.0170 & CORALIE \\
\hline
\end{tabular}
\end{center}
\end{table}

\section{Results and Discussion}
\label{params}

\subsection{Stellar Parameters}

The individual NOT/FIES spectra of WASP-35 were co-added to produce a single spectrum with an average S/N of around 100:1. For WASP-48, the SOPHIE spectra were co-added to produce a spectrum with typical S/N of around 80:1. For WASP-51, the CORALIE spectra were co-added to produce a single spectrum with an average S/N of around 150:1. The standard pipeline reduction products were used in the analysis. 

The analysis was performed using the methods given in \citet{gillon09b}. The H$\alpha$ line was used to determine the effective temperature, \teff, while the Na {\sc i} D and Mg {\sc i} b lines were used as surface gravity, \logg\, diagnostics. The elemental abundances were determined from equivalent width measurements of several clean and unblended lines. Atomic line data was mainly taken from the \citet{kurucz95} compilation, but with updated van der Waals broadening coefficients for lines in \citet{barklem00} and $\log gf$ values from \citet{gonzalez00}, \citet{gonzalez01} or \citet{santos04}. A value for microturbulence, \mictrb, was determined from Fe~{\sc i} using the method of \citet{magain84}. The parameters obtained from the analyses are listed in Table~\ref{tab:params}. The quoted error estimates include that given by the uncertainties in \teff, \logg\ and \mictrb, as well as the scatter due to measurement and atomic data uncertainties.

The projected stellar rotation velocity (\vsini) was determined by fitting the profiles of several unblended Fe~{\sc i} lines. Values for macroturbulence, \mactrb, of 3.3~$\pm$~0.3~\kms, 3.2~$\pm$~0.3~\kms\ and 4.0~$\pm$~0.3~\kms\ were assumed for WASP-35, WASP-48 and WASP-51, respectively, based on the calibration by \cite{bruntt10}, and instrumental FWHM of 0.13~$\pm$~0.01~\AA, 0.15~$\pm$~0.01~\AA\ and 0.11~$\pm$~0.01~\AA, respectively, determined from the telluric lines around 6300~\AA. Best fitting values of \vsini = 3.9 $\pm$ 0.4~\kms, 12.2 $\pm$ 0.7~\kms\ and 3.6 $\pm$ 0.4~\kms, respectively, were obtained.

WASP-51b has been recently announced as HAT-P-30b \citep{johnson11}. A discrepancy between their results and ours is the value of host star metallicity: we find a value of [Fe/H]~=~$-0.08\pm0.08$ whereas \citet{johnson11} report [Fe/H]~=~$0.13\pm0.08$, from an analysis using SME (Spectroscopy Made Easy, \citet{valenti96}). The difference is partly due to a different solar metallicity value adopted in the spectral analysis, where we use log~A(Fe)~=~$7.54\pm0.03$ from \citet{biemont91}, which is 0.04~dex higher than the current \citet{asplund09} solar value of log~A(Fe)~=~$7.50\pm0.04$. We continue to use the \citet{biemont91} value for consistency between analyses rather than adjusting to new values as they are announced. Additionally, we allow the microturbulence value to float and be fitted, finding a best-fit value of 1.3~\kms, in excellent agreement with the calibration in \citet{bruntt10}, compared to the assumed value of 0.85~\kms\ adopted by \citet{valenti05} and used in \citet{johnson11}. Using the lower value of microturbulence would increase our [Fe/H] abundance by 0.07. The slight differences in $T_{\rm eff}$ and log~$g$ values, where \citet{johnson11} find $T_{\rm eff} = 6300\pm90$ and log~$g = 4.36\pm0.03$, result in a difference of a further 0.03~dex. These three factors combine to reduce the discrepancy to only 0.07~dex, which is within the quoted uncertainties.


\begin{table}[h]
\caption{Stellar parameters of WASP-35, WASP-48 and WASP-51 from Spectroscopic Analysis.}
\begin{center}
\begin{tabular}{cccc}
\hline
Parameter & WASP-35 & WASP-48 & WASP-51 \\
\hline
\teff      & 6050 $\pm$ 100 K   & 6000 $\pm$ 150 K & 6250 $\pm$ 100 K \\
\logg      & 4.4 $\pm$ 0.1      & 4.5 $\pm$ 0.15 & 4.3 $\pm$ 0.1  \\
\mictrb    & 1.3 $\pm$ 0.1 \kms & 1.3 $\pm$ 0.2 \kms & 1.3 $\pm$ 0.1 \kms \\
\vsini     & 2.4 $\pm$ 0.6 \kms & 12.2 $\pm$ 0.7 \kms & 3.6 $\pm$ 0.4 \kms \\
 & \\
{[Fe/H]}   &$-$0.15 $\pm$ 0.09  & $-$0.12 $\pm$ 0.12 & $-$0.08 $\pm$ 0.08 \\
{[Na/H]}   &$-$0.16 $\pm$ 0.11  & $-$0.14 $\pm$ 0.07 & $-$0.11 $\pm$ 0.05 \\
{[Mg/H]}   &$-$0.01 $\pm$ 0.11  & $-$0.09 $\pm$ 0.07 & $-$0.08 $\pm$ 0.08 \\
{[Al/H]}   & -                  & ~0.02 $\pm$ 0.14 & - \\
{[Si/H]}   &$-$0.06 $\pm$ 0.05  & ~0.04 $\pm$ 0.08   & $-$0.01 $\pm$ 0.09 \\
{[Ca/H]}   &$-$0.01 $\pm$ 0.12  & $-$0.10 $\pm$ 0.11 & ~0.05 $\pm$ 0.12 \\
{[Sc/H]}   &$-$0.02 $\pm$ 0.07  & ~0.13 $\pm$ 0.09   & ~0.01 $\pm$ 0.06 \\
{[Ti/H]}   &$-$0.03 $\pm$ 0.10  & ~0.02 $\pm$ 0.13   & $-$0.03 $\pm$ 0.08 \\
{[V/H]}    &$-$0.13 $\pm$ 0.06  & $-$0.17 $\pm$ 0.16 & $-$0.10 $\pm$ 0.09 \\
{[Cr/H]}   &$-$0.10 $\pm$ 0.09  & ~0.09 $\pm$ 0.16   & $-$0.01 $\pm$ 0.06 \\
{[Mn/H]}   &$-$0.29 $\pm$ 0.08  & -  & $-$0.27 $\pm$ 0.11 \\
{[Co/H]}   &$-$0.22 $\pm$ 0.10  & -  & - \\
{[Ni/H]}   &$-$0.16 $\pm$ 0.07  & $-$0.18 $\pm$ 0.09 & $-$0.10 $\pm$ 0.06 \\
log A(Li)  &   ~2.35 $\pm$ 0.08  & $<$ 1.3 & ~2.96 $\pm$ 0.08 \\
Mass       &   1.10 $\pm$ 0.08  & 1.09 $\pm$ 0.08 $M_{\sun}$ & 1.20 $\pm$ 0.09 $M_{\sun}$ \\
Radius     &   1.09 $\pm$ 0.14  & 1.09 $\pm$ 0.14 $R_{\sun}$ & 1.28 $\pm$ 0.17 $R_{\sun}$ \\
\hline
\end{tabular}
\end{center}
\label{tab:params}
{\bf Note:} Mass and radius estimates using the \cite{torres10b} calibration.
\end{table}

The lithium abundance in WASP-35 implies an age of $\ga$2~Gyr according to the calibration of \citet{sestito05} while the measured $v \sin i$ gives a rotational period of $P_{\rm rot} \simeq 23 \pm 6.5$~days, assuming $i = 90^{\circ}$, which yields gyrochronological age of $\sim 6^{+5}_{-4}$~Gyr using the relation of \citet{barnes07}. A lack of stellar activity is indicated by the absence of Ca II H+K emission in the spectra.

The non-detection of lithium in the WASP-48 spectrum suggests that the star is several Gyr old, and the lack of any Ca H+K emission is consistent with this. However, the stellar rotation rate of $4.5\pm0.6$~days from the {\vsini} measurement, implies an age of only $\sim 0.2^{+0.2}_{-0.1}$~Gyr using the \citet{barnes07} gyrochronological relation. This discrepency is discussed further below.

The presence of strong lithium absorption in the WASP-51 spectrum suggests that the star is $\la$1~Gyr old. The stellar rotation period of $18.0 \pm 3.1$~days found from the {\vsini} measurement results in an age of $\sim 5.6^{+7.2}_{-2.9}$ Gyr, though this is can be regarded as an upper limit, since the stellar rotation rate may be higher due to the unknown value of stellar inclination ($i$). 

\subsection{System Parameters}

We simultaneously analysed all photometry and radial velocity data using a Markov-Chain Monte Carlo (MCMC) analysis, as set out in \citet{cameron07}, modified to calculate the mass of the host star with a calibration on $T_{\mbox{eff}}$, log $\rho$ and [Fe/H] as described in \citet{enoch10}. The photometry provides the stellar density while the effective temperature and metallicity of the star are obtained through spectral analysis, given above.

We initially allowed the value of the eccentricity of each planetary orbit to float, which resulted in values of $e = 0.057^{+0.064}_{-0.032}$, $e = 0.058^{+0.058}_{-0.035}$ and $e = 0.040^{+0.044}_{-0.028}$ for WASP-35b, WASP-48b and WASP-51b, respectively. The eccentricity of all three orbits are consistent with zero based on the Lucy-Sweeney test \citep{lucy71} on the data for each system, which yields 100\% probability that these eccentricity values could be due to chance if the underlying orbit were circular for all three systems. With no strong evidence for eccentricity in the orbits, we then performed second analyses of all systems, fixing the eccentricities to zero.

\begin{figure}[h!]
\begin{center}
\includegraphics[angle=90,width=125mm,height=80mm]{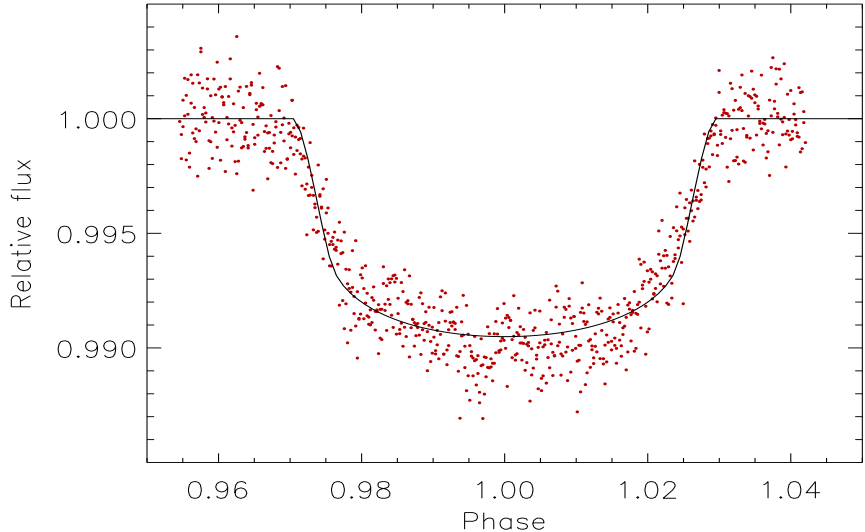}
\caption{RISE photometry shown with the low impact paramter (b=0.4) model; a poor fit compared to the fit with b=0.7 in Figure \ref{fig:wasp48phot}.}
\label{fig:lowimp}
\end{center}
\end{figure}

The MCMC results for WASP-48b reveal a slightly evolved star with a large radius of 1.75~$R_{\odot}$ (discussed below). This radius is around $0.6 R_{\odot}$ greater than expected for that of a star of the same mass on the Main Sequence. To explore this, we performed a further analysis, imposing a Main Sequence constraint on WASP-48, such that a Gaussian prior is used on the stellar radius at each step in the analysis, centred on radius $R_{\ast} = M_{\ast}^{0.8}$, again allowing the eccentricity to float. This forced the stellar radius estimate down to $1.13 R_{\odot}$, artificially inflating the eccentricity estimate to $0.38\pm0.04$ to compensate, since the stellar radius is calculated from the photometry, including a term $\frac{1 + e \mbox{sin} \omega}{\sqrt{1-e^2}}$: an eccentricity of 0.4 produces a stellar radius value 0.65 times that of the estimate with eccentricity at zero. However, the Lucy-Sweeney test result is still 100\%, and there is no evidence for such a high eccentricity in the radial velocity curve, where the $\chi^2$ values are 33.3 and 8.6 for eccentric and circular orbits, respectively. Imposing the Main Sequence constraint with the eccentricity held fixed at zero decreases the stellar radius to $1.31\pm0.09 R_{\odot}$ and decreases the impact parameter from 0.7 to 0.4 to compensate: such a reduction in impact parameter leads directly to a stellar radius estimate 0.70 times smaller. However, the much lower impact parameter estimate produces a model that does not well match the high quality follow up photometric data: compare the fit to ingress and egress in Figure \ref{fig:lowimp}, showing the lower impact parameter model, with that in Figure \ref{fig:wasp48phot}. Since neither analysis with Main Sequence constraint imposed produces realistic models for the photometry and spectroscopy, the results from the analysis with no Main Sequence constraint are clearly preferred.

We also tested the imposition of the Main Sequence constraint on WASP-35 and WASP-51, producing negligible change in all parameters, well within the parameter uncertainties, which implies that the photometry is sufficient to well constrain the results. 

The best-fit parameters for each systems, with eccentricity fixed to zero and no Main Sequence constraint imposed, are given in Table \ref{tab:results}. WASP-35 is found to have a mass of $1.07~\pm~0.02~M_{\odot}$ and radius of $1.09~\pm~0.02~R_{\odot}$, while WASP-35b has a mass of $0.72~\pm~0.06~M_J$ and radius of $1.32~\pm~0.03~R_J$, giving a density of $0.32~\pm~0.03~\rho_J$. For WASP-48, the best-fit model gives a stellar mass of $1.19\pm0.04 M_{\odot}$ and radius of $1.75\pm0.07 R_{\odot}$, giving a stellar density of $0.22\pm0.02 \rho_{\odot}$. WASP-48b is found to have a mass of $0.98\pm0.09 M_J$ and radius of $1.67\pm0.08 R_J$, giving a density of $0.21^{+0.04}_{-0.03} \rho_J$. 

The best-fit result for WASP-51 gave a stellar mass $1.18\pm0.03 M_{\odot}$, stellar radius $1.33\pm0.03 R_{\odot}$, planetary mass $0.76\pm0.05 M_J$, planetary radius $1.42\pm0.04 R_J$, giving a density of $0.26\pm0.02 \rho_J$. Under the name HAT-P-30b, based on HAT-5 (Arizona) and HAT-9 (Hawaii) photometry of a total of around 3,200 good datapoints, follow-up high quality photometric observations with Kepler-Cam on the FLWO 1.2m telescope, and Keck and Subaru radial velocity measurements, \citet{johnson11} report a planetary mass of $0.71\pm0.03 M_J$, and radius of $1.34\pm0.07 R_J$, which agree with our values.

\begin{sidewaystable}
\begin{center}
\caption{System Parameters from simultaneous MCMC analysis of all photometric and spectroscopic data.}
\label{tab:results}
\begin{tabular}{llrrr}
\hline
Parameter & Symbol & WASP-35 & WASP-48 & WASP-51  \\
\hline
Period (days) & $P$ &                        $3.161575~\pm~0.000002$  & $2.143634~\pm~0.000003$ & $2.810603~\pm~0.000008$ \\
Transit Epoch (HJD) & $T_0$ &                $5531.47907~\pm~0.00015$ & $5364.55043~\pm~0.00027$  & $5571.70057~\pm~0.00020$ \\
Transit duration (days) & $D$ &              $0.1278\pm0.0005$  & $0.1327\pm0.0013$  & $0.0920\pm0.0008$ \\
Planet/Star area ratio & $R_p^2/R_{\ast}^2$ & $0.0154~\pm~0.0001$  & $0.0096~\pm~0.0002$  & $0.0122~\pm~0.0002$ \\
Impact Parameter & $b$ &                      $0.30\pm0.04$ & $0.73\pm0.03$ & $0.87\pm0.01$ \\
Stellar Reflex Velocity (ms$^{-1}$) & $K_1$ & $94.82^{+7.28}_{-6.92}$ & $136.02^{+11.07}_{-11.44}$ & $96.70^{+6.07}_{-6.04}$ \\
Centre-of-mass Velocity (kms$^{-1}$) & $\gamma$ & $17.72~\pm~0.01$  & $-19.683~\pm~0.001$ & $44.6768~\pm~0.0001$ \\
Orbital separation (AU) & $a$ &                 $0.04317~\pm~0.00028$ & $0.03444~\pm~0.00043$ & $0.04118~\pm~0.00031$ \\
Orbital inclination (deg) & $i$ &               $87.96^{+0.31}_{-0.25}$ & $80.09^{+0.69}_{-0.55}$  & $82.48^{+0.14}_{-0.17}$ \\
Orbital eccentricity & $e$ &                    0 (fixed) & 0 (fixed) & 0 (fixed) \\
Orbital distance / radius ratio & $a / R_{\ast}$ & $8.53\pm0.19$  & $4.23^{+0.24}_{-0.19}$ & $6.67\pm0.17$\\
Stellar mass (M$_{\odot}$) & $M_{\ast}$ &          $1.07\pm0.02$  & $1.19~\pm~0.04$  & $1.18\pm0.03$ \\
Stellar radius (R$_{\odot}$) & $R_{\ast}$ &        $1.09\pm0.02$ & $1.75\pm0.07$  & $1.33\pm0.03$ \\
Stellar surface gravity (log $g_{\odot}$) & log $g_{\ast}$ & $4.40\pm0.01$  & $4.03\pm0.03$  & $4.26\pm0.01$ \\
Stellar density ($\rho_{\odot}$) & $\rho_{\ast}$ &           $0.83\pm0.03$ & $0.22\pm0.02$  & $0.50\pm0.02$ \\
Stellar metallicity & $[$Fe/H$]$ &                           $-0.15\pm0.09$  & $-0.12\pm0.12$  & $-0.08\pm0.08$ \\
Stellar effective temperature & T$_{\mbox{eff}}$ &           $5990~\pm~80$ & $5920~\pm~150$ & $6250~\pm~100$ \\
Planet mass (M$_J$) & $M_p$ &                          $0.72\pm0.06$ & $0.98\pm0.09$  & $0.76\pm0.05$ \\
Planet radius (R$_J$) & $R_p$ &                        $1.32\pm0.03$  & $1.67\pm0.08$ & $1.42\pm0.04$ \\
Planet surface gravity (log $g_J$) & log $g_p$ &       $2.98\pm0.04$ & $4.03\pm0.03$ & $2.93\pm0.03$ \\
Planet density ($\rho_J$) & $\rho_p$ &                 $0.32\pm0.03$ & $0.21\pm0.04$  & $0.26\pm0.02$ \\
Planet temperature (A=0, F=1) (K) & T$_{\mbox{eq}}$ &  $1450~\pm~20$ & $2030~\pm~70$ & $1710~\pm~30$ \\
\hline
\end{tabular}
\end{center}
\end{sidewaystable}

\subsection{Isochrone Analysis}

\begin{figure}[h!]
\begin{center}
\includegraphics[angle=270,width=80mm]{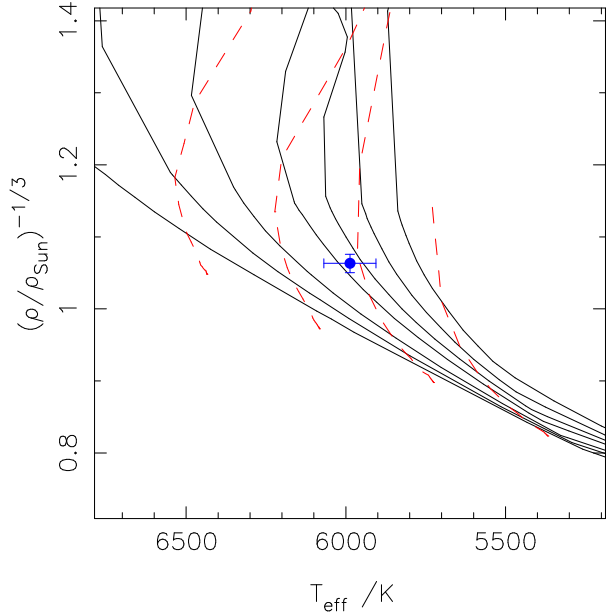}
\caption{Isochrone tracks from \citet{marigo08} for WASP-35. Isochrones (solid lines, left to right) are: 1.0, 1.99, 3.16, 5.01, 6.30, 7.94, 10.0 Gyr. Evolutionary tracks (dashed lines) are for 1.2, 1.1, 1.0, 0.9 $M_{\odot}$ stars.}
\label{fig:35iso}
\end{center}
\end{figure}

\begin{figure}[h!]
\begin{center}
\includegraphics[angle=270,width=80mm]{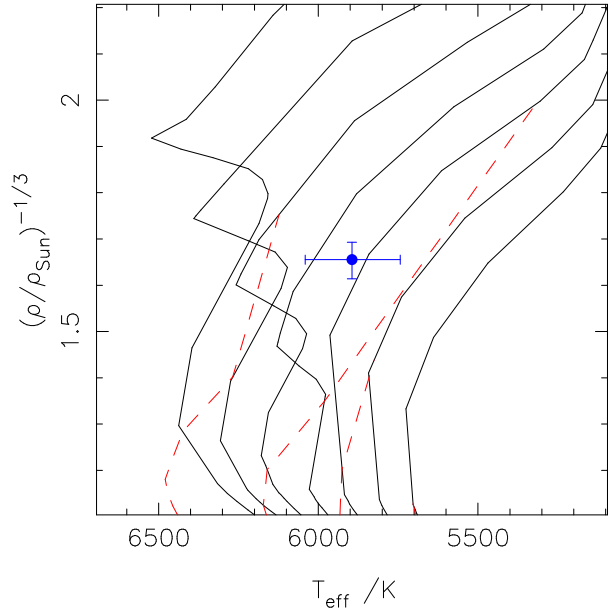}
\caption{Isochrone tracks from \citet{marigo08} for the slightly evolved WASP-48 ($\sim 8$~Gyr, $1.75 R_{\odot}$). Isochrones (solid lines, left to right) are 3.16, 3.98, 5.01, 6.30, 7.94, 10.0, 12.5 Gyr. Evolutionary tracks (dashed lines) are for 1.2, 1.1, 1.0, 0.9 $M_{\odot}$ stars.}
\label{fig:48iso}
\end{center}
\end{figure}

\begin{figure}[h!]
\begin{center}
\includegraphics[angle=270,width=80mm]{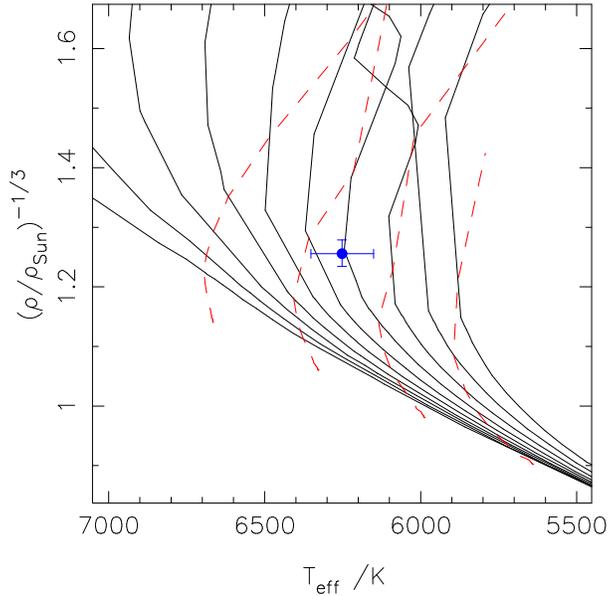}
\caption{Isochrone tracks from \citet{marigo08} for WASP-51. Isochrones (solid lines, left to right) are:  1.0, 1.25, 1.58, 1.99, 2.51, 3.16, 3.98, 5.01, 6.30, 7.94 Gyr. Evolutionary tracks (dashed lines) shown are for 1.3, 1.2, 1.1, 1.0 $M_{\odot}$ stars.}
\label{fig:51iso}
\end{center}
\end{figure}

The values for stellar density, effective temperature and metallicity from the MCMC analyses (MS constraint off, eccentricity fixed to zero) were used in an interpolation of the Padova stellar evolution tracks \citep{girardi02,marigo08}, shown in Figures \ref{fig:35iso}, \ref{fig:48iso} and \ref{fig:51iso}. This resulted in an age for WASP-35 of $5.01~\pm~1.16$~Gyr and a mass of $1.03~\pm~0.04 M_{\odot}$. 

WASP-48 was found to have an age of $7.9^{+2.0}_{-1.6}$~Gyr, and mass of $1.08\pm0.06 M_{\odot}$, supporting the results from the MCMC analyses and lack of lithium and Ca H+K that WASP-48 is an old, slightly evolved star. However, the inferred age from the stellar rotation period is still only $0.6^{+0.4}_{-0.2}$~Gyr when based on a rotational period of $7.2\pm0.5$~days found from the larger stellar radius of $1.75\pm0.07 R_{\odot}$ (with \vsini\~=~$12.2\pm0.7$~\kms, as before). This apparent contradiction may be due to stellar spin-up by the planet. The close orbits of most known transiting exoplanets produce strong tidal forces between the planet and host star, which lead eventually to orbital circularisation, synchronisation and decay. There is evidence for tidal circularisation among the known transiting exoplanets, seen in the period-eccentricity relation where planets with very short orbital periods have circular orbits while those with longer periods show a range of eccentricities, but the synchronisation of stellar rotation is generally ruled out for most known planet-star systems, with timescales orders of magnitude larger than the Hubble time \citep{mazeh08,pont09}. A few exceptions exist where the companion is massive, e.g. the unique case of $\tau$~Boo for which the planetary orbital period coincides closely with the mean rotation period of the stellar differential rotation pattern determined from the Zeeman-Doppler imaging \citep{fares09}. WASP-48b orbits with a period of 2.14~days (a~=~0.034~AU), and while it is not a massive companion, the evolved nature of the star giving a radius of $1.75~R_{\odot}$ results in a relatively short stellar tidal synchronisation time of $11^{+3}_{-2}$~Gyr. A full tidal analysis will be discussed in the forthcoming paper of \citet{brown11}. 

For WASP-51, the interpolation yields an age of $4.0^{+1.0}_{-0.8}$~Gyr, and mass of $1.16\pm0.04 M_{\odot}$. 

An alternative analysis was also performed using the Yonsei-Yale isochrones \citep{demarque04}, which produced an age estimate of $4^{+2}_{-1}$~Gyr and a mass estimate of $1.04~\pm~0.04 M_{\odot}$ for WASP-35, $6.0^{+2.0}_{-1.0}$~Gyr and $1.14^{+0.06}_{-0.07} M_{\odot}$ for WASP-48 and $3.0^{+1.0}_{-0.5}$~Gyr and $1.20^{+0.03}_{-0.05} M_{\odot}$ for WASP-51.

The results of both isochrone analyses are consistent within uncertainties, though the Padova interpolations yield higher ages for each system than the Yonsei-Yale interpolations, and mass values agree with the calibrated values from the MCMC analysis of $1.08\pm0.03 M_{\odot}$, $1.19\pm0.04 M_{\odot}$ and $1.18\pm0.02 M_{\odot}$ for WASP-35, WASP-48 and WASP-51, respectively.

\section{Summary}

We have presented observations of WASP-35b, WASP-48b and WASP-51b, new transiting exoplanets of 0.72~$M_J$ and 1.32~$R_J$, 0.98~$M_J$ and 1.67~$R_J$, and 0.76~$M_J$ and 1.42~$R_J$, respectively. We have obtained photometry from both WASP-North and -South instruments, as well as high-quality follow-up photometry from RISE, FTS and TRAPPIST. Radial velocity measurements from the FIES, SOPHIE and CORALIE instruments allowed us to confirm the planetary nature of the companions. 

WASP-48b may have spun-up its slightly evolved host star, which shows a high \vsini\ value, otherwise contradicting the estimated stellar age of around 8~Gyr. WASP-51b has been recently announced as HAT-P-30b, with mass $0.71 M_J$ and radius $1.34 R_J$, values that agree with our results, within uncertainties.


\acknowledgments
\small
WASP-North is hosted by the Isaac Newton Group on La Palma and WASP-South is hosted by the South African Astronomical Observatory (SAAO) and we are grateful for their ongoing support and assistance. Funding for WASP comes from consortium universities and from the UK`s Science and Technology Facilities Council.
The RISE instrument mounted on the Liverpool Telescope was designed and built with resources made available from Queen's University Belfast, Liverpool John Moores University and the University of Manchester. The Liverpool Telescope is operated on the island of La Palma by Liverpool John Moores University in the Spanish Observatorio del Roque de los Muchachos of the Instituto de Astrofisica de Canarias with financial support from the UK Science and Technology Facilities Council. We thank Tom Marsh for the use of the ULTRACAM pipeline.
TRAPPIST is a project funded by the Belgian Fund for Scientific Research (FNRS) under the grant FRFC 2.5.594.09.F, with the participation of the Swiss National Science Fundation (SNF).  M. Gillon and E. Jehin are FNRS Research Associates.
The research leading to these results has received funding from the European Community`s Seventh Framework Programme (FP7/2007-2013) under grant agreement number RG226604 (OPTICON).
\normalsize

\bibliographystyle{aa}
\bibliography{wasp35_48_51.bib}

\end{document}